\documentclass[prl,twocolumn]{revtex4}
\pdfoutput=1
\usepackage{amsfonts,amssymb,amsmath}            
\usepackage[]{graphics,graphicx,epsfig}            
\usepackage{amsthm}
\def\identity{\leavevmode\hbox{\small1\kern-3.8pt\normalsize1}}

\newcommand{\ket}[1]{\left | #1 \right\rangle}
\newcommand{\bra}[1]{\left \langle #1 \right |}
\newcommand{\half}{\mbox{$\textstyle \frac{1}{2}$}}

\newcommand{\proj}[1]{\ket{#1}\bra{#1}}
\newcommand{\Tr}{\text{Tr}}
\renewcommand{\epsilon}{\varepsilon}
\begin{document}

\title{Computation on Spin Chains with Limited Access}
\date{\today}

\author{Alastair \surname{Kay}}
\affiliation{Centre for Quantum Computation,
             DAMTP,
             Centre for Mathematical Sciences,
             University of Cambridge,
             Wilberforce Road,
             Cambridge CB3 0WA, UK}
\author{Peter J.~\surname{Pemberton-Ross}}
\affiliation{Centre for Quantum Computation,
             DAMTP,
             Centre for Mathematical Sciences,
             University of Cambridge,
             Wilberforce Road,
             Cambridge CB3 0WA, UK}

\begin{abstract}
We show how to implement quantum computation on a system with an intrinsic Hamiltonian by controlling a limited subset of spins. Our primary result is an efficient control sequence on a nearest-neighbor $XY$ spin chain through control of a single site and its interaction with its neighbor. Control of an array of sites yields sufficient parallelism for the implementation of fault-tolerant circuits. The framework exposes contradictions between the control theoretic concept of controllability with the ability of a system to perform quantum computation.
\end{abstract}

\maketitle

{\em Introduction:} What does it take to implement a quantum computation in a given physical system? This would seem to be a fundamental question, for which a sufficient set of conditions is well known \cite{criteria}; implementation of single qubit rotations on any spin, and a nearest-neighbor two-qubit gate. However, since that degree of control seems to be extremely demanding, it is vital to understand how little control is required. In fact, there are some systems whose internal dynamics are sufficient to implement computations \cite{karl,kay:08}. However, these have to be carefully designed, and still require the ability to prepare the initial (product) state. On the other hand, reintroducing control over a single spin in principle gives sufficient control for almost all Hamiltonians \cite{uqi}. This architecture is an attractive proposition for some experimental implementations. For instance, while it has been shown that full control over a pair of superconducting qubits can be achieved, the physical layout of such devices is highly asymmetric \cite{yamamoto}, and is not easily scaled up. Thus, we might be able to consider manufacturing a uniform system with some fixed interaction, and can concentrate all our design efforts in generating controllability at the end of a chain, which is allowed to be non-symmetric.

The proofs of controllability of these interface schemes \cite{uqi} make no claims regarding efficiency. Some examples of Hamiltonians have been specifically constructed to allow efficient control sequences \cite{kay:08,kay:09}. While much less complicated than those of \cite{karl, kay:08} which function without any control, they are still unrealistic. In this paper, we develop efficient analytic control sequences for a much more natural class of Hamiltonians; spin chains. The main ingredients are an encoding of information in the diagonal basis of the Hamiltonian, and the use of Rabi oscillations to induce transitions between these states.

{\em Generic Controllability:} Consider an $N$-qubit Hamiltonian $H$, with control field $h_1$. Each arbitrarily entangled eigenvector $\ket{\lambda_x}$ can be identified with a logical basis state $\ket{x_L}$, $x\in\{0,1\}^N$. 
Generically, the eigenvalues $|\lambda_x|$ and differences $|\lambda_x-\lambda_y|$ are unique, and $\bra{\lambda_x}h_1\ket{\lambda_y}\neq 0$. Under these assumptions, the field
$$
h_X^n=B\!\!\sum_{\stackrel{x\in\{0,1\}^N}{x_n=0}}\frac{1}{\bra{\lambda_{x\oplus n}}h_1\ket{\lambda_x}}\cos\left((\lambda_x-\lambda_{x\oplus n})t\right)h_1
$$
applies the logical $X$ rotation on qubit $n$ (up to some phases, which we consider later). $x\oplus n$ is used to denote the flipping of bit $n$ in the string $x$. Naturally, $h_X^n$ only makes sense if $\bra{\lambda_{x\oplus n}}h_1\ket{\lambda_x}\neq 0$. Due to the assumed uniqueness of gaps, each term in the sum is on resonance with a single transition so that, by applying the rotating wave approximation (RWA, which requires that the detunings of different energy gaps is much greater than $B$), the effective Hamiltonian in the interaction picture is
$$
H_{\text{eff}}=B\!\!\sum_{\stackrel{x\in\{0,1\}^N}{x_n=0}}\ket{\lambda_x}\bra{\lambda_{x\oplus n}}+\ket{\lambda_{x\oplus n}}\bra{\lambda_x},
$$
which evidently provides the logical $X$ rotation that we desire, by any angle $Bt_X$. The effect of returning to the Schr\"odinger picture is that this rotation is followed up by $\sum_xe^{-i\lambda_x t_X}\proj{\lambda_x}$. A cNOT gate (up to an identical phase condition) is implemented in a similar fashion,
$$
h_{cNOT}^{n,m}=B\!\!\!\!\sum_{\stackrel{x\in\{0,1\}^N}{x_m=1,x_n=0}}\frac{1}{\bra{\lambda_{x\oplus n}}h_1\ket{\lambda_x}}\cos\left((\lambda_x-\lambda_{x\oplus n})t\right)h_1
$$
with control qubit $m$ and target $n$. In order to have full controllability, we just need to demonstrate how to implement arbitrary $Z$ rotations on any spin, $n$. This can also be used to cancel the phases that accrue due to the interaction picture. The first step is to negate the effect of the phases when implementing an identity operation, using the standard NMR technique of refocusing -- by applying the cyclic permutation 
$\sum_{x}\ket{\lambda_{x+1\mod 2^N}}\bra{\lambda_x}$ 
$2^N$ times, waiting the same time $t_Z$ between each application, then all eigenvectors accumulate the same phase, $t_Z\Tr(H)$. Independently varying the waiting times in different intermediate states allows different phases to be applied to different eigenvectors, which is precisely what we need, thereby proving controllability of a generic Hamiltonian.

This technique is, in the majority of cases, wildly inefficient, for several reasons. Primarily, since there is an exponential number of eigenvectors, $B$ must be exponentially small if the control field is to be bounded, so gates take exponentially long. Equally, to cycle through all the eigenvectors for the phase gate is an exponential process. While these techniques are not necessarily unique, the on-resonant control would seem to be an essential component of any such scheme. How can any scheme be efficient? Introducing degeneracies into the system reduces the number of terms that we sum over. However, care is required since, if we have that $\lambda_x-\lambda_{x\oplus n}$ is independent of $x$ (such that $h_X^n$ is only a single term), then there is too much degeneracy for $h_{cNOT}$, and the existing proof of controllability breaks down. Bizarrely, to get efficient computation, we have to make it harder to prove controllability! Worthy of emphasis is that controllability typically applies to the control of the entire Hilbert space, whereas efficient quantum computation only requires control over a subsystem.

{\em Computation on spin chains:} While it may be interesting to understand generic systems, the Hamiltonians that are accessible in the laboratory are far from generic, so the preceding arguments need not apply. We shall now show how to use the basic ideas introduced to efficiently compute on a spin chain of the form
\begin{equation}
H=\half\sum_{n=1}^NJ_n((1+\gamma)XX+(1-\gamma)YY)_{n,n+1}-\half\sum_{n=1}^NB_nZ_n.	\label{eqn:ham}
\end{equation}
This Hamiltonian is exactly solvable \cite{lieb}, the first step being to perform the Jordan-Wigner transformation $a_n^\dagger=\sigma_n^+\prod_{m=1}^{n-1}Z_m$. These can then be transformed into a set of non-interacting fermions,
$$
H_f=\sum_{n=1}^N\lambda_nb_n^\dagger b_n.
$$
via a Bogoliubov transformation and diagonalization of an $N\times N$ tridiagonal matrix.
We shall assume that the coupling strengths $J_n$ are known, although they can be identified experimentally \cite{burgarth:08} after preparing the system in some initial state \cite{burgarth:07}. For pedagogical reasons, we introduce three control fields, which only act on the first two spins \footnote{A single field, such as $h_1+h_2+h_3$, is typically sufficient.},
\begin{eqnarray}
h_1&=&X_1	\nonumber\\
h_2&=&\half((1+\gamma)XX+(1-\gamma)YY)_{1,2}	\nonumber\\
h_3&=&Z_1Z_2.	\nonumber
\end{eqnarray}
These can be transformed into the $\{a_n\}$ or $\{b_n\}$ basis. We can assume $J_1=B_1=0$, which means that $b_1=a_1$ and $\lambda_1=0$. In cases such as $\gamma=0$, it is already guaranteed that the $\{\lambda_n\}$ are unique and that $\alpha_n:=\bra{0}b_1h_2b_n^\dagger\ket{0}\neq 0$ \cite{kay:review}, where $\ket{0}$ denotes the vacuum (i.e.~ground) state of the system. The uniqueness of the $\{\lambda_n\}$ is sufficient to give the condition of uniqueness of the $\{|\lambda_n|\}$ since we could tune the field $B_1$ which, working in an offset system where we keep $\lambda_1=0$, rescales all other eigenvalues by $B_1$, sufficient to move them off any degeneracies due to the existence of $\pm\lambda_n$ eigenvalue pairs. It is also sufficient to ensure that none of the eigenvalues are exponentially small. Henceforth, we assume these conditions hold.

Instead of proving universal computation on the full Hilbert space, we shall just consider a subspace where the logical qubits are described by pairs of fermions. The initial state is of the form 
$$
\ket{0_L}^{\otimes\lfloor (N-1)/2\rfloor}=\prod_{m=1}^{\lfloor (N-1)/2\rfloor}b_{2m}^\dagger\ket{0},
$$
and the raising operator for the $n^{th}$ logical qubit is $\sigma_n^+=b_{2n+1}^\dagger b_{2n}$. The primary reason for this choice is that if we were to encode in single fermion states, then when moving states around the lattice, they generate exchange phases, which correspond to controlled-phase gates. Encoding in a $\ket{01}_L,\ket{10}_L$ subspace negates these effects \cite{kay-2006b}. Note that $b_1$ is not used to encode a qubit, and is instead kept free, as workspace.

\begin{figure}[!tb]
\begin{center}
\includegraphics[width=0.45\textwidth]{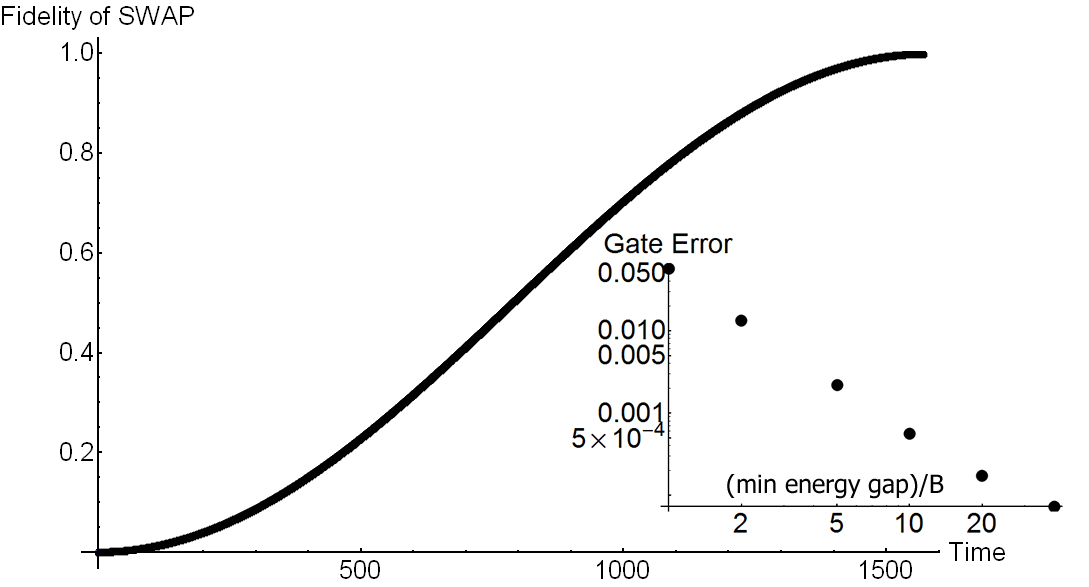}
\end{center}
\vspace{-0.5cm}
\caption{The effectiveness of the swap gate between the $b_1$ mode and the $b_n$ with minimum eigenvalue for $N=101$ and the coupling scheme of Eq.~(\ref{eqn:new_couple}), offset by using $B_1=\sqrt{2}$. The inset shows how decreasing $B$ increases the accuracy. The fidelity is calculated as the overlap between the evolution of an initial state $\ket{0}\ket{\lambda_{100}}$ and the target state $\ket{1}\ket{0}^{\otimes 100}$. All quantities are dimensionless by taking $\hbar=1$.}\label{fig:swap}
\vspace{-0.5cm}
\end{figure}

All protocols in the computation require the field 
$$B_n(t)=B\cos(\lambda_nt)h_2,$$ 
which implements the effective Hamiltonian
$$
\half \alpha_nB(b_1^\dagger b_n+b_n^\dagger b_1).
$$
The corresponding unitary evolution is just a swap between the two modes $b_1$ and $b_n$, except that when we reassert the normal ordering of the fermionic modes, a phase factor of $\prod_{m=2}^{n-1}(2b_m^\dagger b_m-\identity)$ arises if the swap occurred. The sequence of $\prod_{m=2}^{n-1}(2b_m^\dagger b_m-\identity)$ is precisely the c-phase gates mentioned previously, whose effects are negated by the encoding -- that term calculates the parity of the number of fermions in modes 2 to $n-1$ if there's only 1 fermion in modes 1 or $n$, and this number is fixed due to our encoding. Thus, up to a diagonal gate, $B_n(t)$ can be used to implement a swap of a fermion in mode $n$ onto spin 1. When implementing this swap, one of the two states will always be empty, so the diagonal gate is only a local phase gate, which we will later see how to correct (either we swap a fermion onto the empty state on site 1, or we undo that swap). Once we have implemented $B_{2n}(\pi/(B\alpha_{2n}))$ to swap fermion $2n$ to the first site, we can implement $B_{2n+1}(2\theta/(B\alpha_{2n+1}))$ before applying $B_{2n}(\pi/(B\alpha_{2n}))$. This returns the fermions to their original positions but the logical qubit $n$ has undergone an $X$-rotation of angle $\theta$, up to the phase gates due to the transformation between the interaction and Schr\"odinger pictures. This protocol also allows the preparation of any eigenstate of $H_f$ and measurement of any logical qubit; swapping $n$ to 1, measuring and swapping back projects the system into a Fock state of $b_n$, and $h_1$ allows the bit to be flipped after measurement. Fig.~\ref{fig:swap} demonstrates the simple swapping protocol for a chain of 101 spins in the single fermion subspace.

A refocusing technique can now be used to perform arbitrary $Z$ rotations. If an $X$ rotation is performed on each logical qubit every $t_Z$, then they each acquire a global phase of the form $(\lambda_{2n}+\lambda_{2n+1})mt_Z$ at times $2mt_Z$. By performing the gate $X_n$ at times $t_Z'$ (instead of $t_Z$) and $2t_Z$, we get a phase rotation of $2(\lambda_{2n+1}-\lambda_{2n})(t_Z-t_Z')$. 

To entangle logical qubits $m$ and $n$, we apply $B_{2m}(\pi/(B\alpha_{2m}))$, swapping the fermionic mode $2m$ onto spin 1, followed by a field 
$B'_n(t)=2B'\cos((\lambda_{2n}-\lambda_{2n+1})t)h_3$ 
for time $\theta/(2B')$, which gives an effective interaction between modes $2n$ and $2n+1$ (in the interaction picture), dependent on the presence or absence of a fermion on the first spin,
\begin{eqnarray}
H_{\text{eff}}&=&B'\alpha_{2n}\alpha_{2n+1}(2b_1^\dagger b_1-\identity)(b_{2n}^\dagger b_{2n+1}+b_{2n+1}^\dagger b_{2n}).	\nonumber
\end{eqnarray}
By applying $B_{2m}(\pi/(B\alpha_{2m}))$, the sequence is completed. The ultimate result is a c-$X$ rotation of angle $\theta$, targeting qubit $n$, up to local rotations.

This proves the possibility of implementing computational gates on a sufficiently large subspace. However, it is not sufficient for efficiency since the timing condition is based on the requirement that $B$ and $B'$ are sufficiently small. This gives two conditions to satisfy, $B\lesssim1$ and $B\max\alpha_n\ll\min|\lambda_n-\lambda_m|$. The first of these arises from the desire to only use finite field strengths. If any eigenvalues, or their gaps, are exponentially small, or overlaps of eigenvectors on the second spin are exponentially small (any of which can happen, albeit rarely), then the gate time must be exponentially long. This loss of practical controllability as a theoretically controllable system closely approaches a symmetric uncontrollable system has recently been identified in \cite{tannor}. In the case of the uniformly coupled chain ($\gamma=0, J_n=1, B_n=0$), the detunings are of the order of $1/N^2$, so gate times are $O(N^2)$. Superior schemes can be designed, such as that introduced in \cite{kay:review}, with $\gamma=B_n=0,$
\begin{equation}
J_{n+1}^2=\frac{3n^2((N-1)^2-n^2)}{N(N-2)(2n-1)(2n+1)}.	\label{eqn:new_couple}
\end{equation}
It has a spectrum with regular spacings of $2/(N-2)$ and $\alpha_n=1/\sqrt{N-1}$, meaning that it can implement gates in a time $O(N)$, which is optimal if $J\sim O(1)$. Some care has to be taken with the two-qubit gate, since the gaps between eigenvalues are highly degenerate. The first step in overcoming this is to make a suitable association between the numbering of the fermionic modes and their eigenvalues, $\lambda_{2n}=-1+2(n-1)/(N-2), \lambda_{2n+1}=(2n-1)/(N-2)$, which means that applying $B'_n(t)$ uses a frequency greater than half of the total energy range i.e.~each mode can only couple to one other mode, and corresponds to applying the same $X$ rotation on every logical qubit simultaneously, if the $b_1$ mode is occupied. To localixe this effect on a single target, we apply a logical $Z$ gate at the start, and half way through the evolution, on the qubits where we don't what the cNOT applied. Since $Z\sqrt{X}Z\sqrt{X}=\identity$, the evolution is canceled, and if the $b_1$ mode was not occupied, we only get $ZZ=\identity$.

The engineered coupling scheme of Eq.~(\ref{eqn:new_couple}) is particularly amenable to the final step of the analysis -- an estimation of how the gate error scales with $N$ and $B$. Instead of considering $B_n(t)$, we will replace it with
\begin{eqnarray}
B_n(t)&=&\frac{B}{4}\cos(\lambda_nt)((1+\gamma)XX+(1-\gamma)YY)_{1,2}	\nonumber\\
&&+\frac{B}{4}i\sin(\lambda_nt)((1+\gamma)XY-(1-\gamma)YX)_{1,2},	\nonumber
\end{eqnarray}
reducing our reliance on the RWA,
$$
H_{\text{eff}}=\sum_{m=1}^N\lambda_mb_m^\dagger b_m+\frac{B}{2\sqrt{N-1}}\sum_{m=2}^Ne^{i\lambda_nt}b_m^\dagger b_1+e^{-i\lambda_nt}b_1^\dagger b_m,
$$
neglecting, for convenience, the string of operators $\prod_{k=2}^{m-1}(2b_k^\dagger b_k-\identity)$ which we know to be irrelevant due to our choice of encoding. 
A suitable rotating basis can be chosen to entirely remove the time dependence. As a first step, we estimate the error in the rotation within the $\{b_1,b_n\}$ subspace by adiabatically eliminating the other levels. This leads to an error of $\varepsilon\sim B^2N\log^2 N$, using
$$
\left|\sum_{\stackrel{m=2}{m\neq n}}^N\frac{1}{\lambda_m-\lambda_n}\right|\leq\sum_{m=1}^{N-1}\frac{N-2}{2m}\sim N\log N.
$$
We also need to estimate the leakage out of this subspace, which can be achieved by assuming the desired evolution of the subspace, in particular the amplitude of the $b_1^\dagger$ mode can be taken to be $\cos(Bt/\sqrt{N})$. Using this, the evolution of the other modes can be solved exactly, and their maximum amplitudes can be bounded. Summing all these reveals a maximum error of $\varepsilon\sim B^2N\log^2 N$. Therefore, by selecting $B^{-1}\sim\sqrt{N}\log(N)$, the error is held constant, and the gate time scales as $O(N\log N)$. In Fig.~\ref{fig:swap}, the error $\varepsilon$ is evaluated numerically for fixed $N$, and indicates $\varepsilon\sim B^{1.9}$. Other coupling schemes are more strongly affected, but a choice of $B\sim\min(\alpha_n)\min|\lambda_m-\lambda_n|/N$ ensures a constant error with increasing $N$.

One might ask how robust this scheme is to fluctuations in the control fields. Since we are using Rabi oscillations, there is a lot of built-in tolerance -- the pulse sequence can be anything provided it has the correct Fourier component with the correct amplitude. Other Fourier components are irrelevant provided they are sufficiently far from the other energy gaps of our system. If the (integrated) amplitude of our Fourier component is slightly wrong, then that means the angle of the implemented $X$ rotation is incorrect by the same fraction, but it is exactly the same possibility of error that all non-topological schemes suffer from. Similarly, if there is a small (compared to $B\alpha_n$) frequency discrepancy, this introduces a small $Z$-component to the $X$ rotation. The Fourier decomposition of control sequences also indicates a link with \cite{pete} where control of the single excitation subspace was demonstrated. Evidently, our fields $B_n(t)$ give efficient controllability of this subspace for any spin preserving network, via Givens rotations (the exchange phases never manifest in the single excitation subspace). In \cite{pete}, the numerical techniques which suggested efficiency were based on a simple on/off switching of $h_2$, which can be directly related to our result by examining the Fourier modes of the square wave.

{\em Fault tolerance:} This interface scheme has many advantages such as not needing to perfectly engineer the system to within tight constraints. Instead, system tomography can feed back into the control sequences. Also, at least theoretically, the majority of the system can be isolated from the environment, thereby decreasing decoherence. Nevertheless, the possibility of error correction remains a concern. This introduces a significant problem to the interface scheme; as the system size increases, the errors accumulate more rapidly than they can be corrected. However, the architecture described here readily generalizes to structures with sufficient parallelism for fault-tolerance \cite{uqi}. Consider the system of Eq.~(\ref{eqn:ham}), but where we control some fixed set of spins $\{k_i\}$, by which we mean that we control the spins $k_i$ and $h_2$ and $h_3$ couplings between neighboring pairs $(k_i-1,k_i)$ and $(k_i,k_i+1)$. By considering the scenario where all these couplings are switched off, the basis defines the computational basis. On each site, if we only ever allow one of the couplings $(k_i-1,k_i)$ or $(k_i,k_i+1)$ to be active at a time, gates can be implemented in time $O(1)$ within a block, or between neighboring blocks. This is sufficient to design a fault-tolerant scheme \cite{gottesman}, although care is required since errors that occur independently on each physical qubit correspond to correlated errors in the encoded basis, constrained within a specific block. The constant sized blocks can be arranged into any geometry, allowing improvements in the fault-tolerant threshold. One would expect a threshold for per spin error rates of the order of $\varepsilon_c/K$ where $\varepsilon_c$ is any fault-tolerant threshold constrained by a locality condition, and $K$ is the number of spins in any given block.

{\em Conclusions:} Simple systems of non-interacting fermions, which can be converted to a wide variety of spin models, including $XX$ and transverse Ising, can be efficiently controlled through the coupling of a single spin to its neighbor, enabling implementation of quantum computation. Without the additional coupling, the structure of the Hilbert space is entirely described by representations of SU($N$), which can be simulated in polynomial time on a logarithmic number of qubits, but introduction of a single controlling interaction breaks this symmetry and potentially permits a computation. The remarkable aspect is the ability to present analytic, efficient, pulse sequences to achieve a computation. We have further discussed how the result generalizes to an array of controllers, which are sufficient to allow a fault-tolerant implementation; a feature absent from previous constructions \cite{kay:08,kay:09}. Our formalism motivates the expectation that most systems, while controllable, cannot be efficiently manipulated. This includes many interesting systems such as Heisenberg chains.

In parallel to this work, Burgarth {\em et al.} have considered the same problem \cite{burgarth:09}. In essence, our work proves when good solutions exist, at which point \cite{burgarth:09} can be used to numerically find control sequences with smaller overheads (no proof for the existence of, or efficient convergence to, solutions is given in \cite{burgarth:09}).

{\em Acknowledgments:}  PJP is supported by an EPSRC Project Studentship.  This research was supported in part by the National Science Foundation under Grant No. PHY05-51164. ASK is supported by Clare College, Cambridge and thanks D.~Burgarth for useful feedback.

\end{document}